\documentclass[useAMS,usenatbib]{mn2e}

\usepackage{graphicx}
\usepackage{amsmath}

\title[A Broadband Scalar Vortex Coronagraph]{A Broadband Scalar Vortex Coronagraph}
\author[R. Errmann, S. Minardi and T. Pertsch]{R. Errmann$^{1,3}$, S. Minardi$^{2}$\thanks{Corresponding author: stefano.minardi@uni-jena.de}, and T. Pertsch$^{2}$\\
$^{1}$Astrophysical Institute and University Observatory, Friedrich-Schiller-Universit\"at Jena, Schillerg\"a{\ss}chen 2-3, 07745 Jena, Germany\\
$^{2}$Institute of Applied Physics, Abbe Center of Photonics, Friedrich-Schiller-Universit\"at Jena, Max-Wien-Platz 1, 07743 Jena, Germany\\
$^{3}$Abbe Center of Photonics, Max-Wien-Platz 1, 07743 Jena, Germany}
\begin{document}

\date{Accepted 2013 July 16.  Received 2013 July 16; in original form 2013 June 7}

\pagerange{\pageref{firstpage}--\pageref{lastpage}} \pubyear{2013}

\maketitle

\label{firstpage}

\begin{abstract}
Broadband coronagraphy with deep nulling and small inner working angle has the potential of delivering images and spectra of exoplanets and other faint objects.
In recent years, many coronagraphic schemes have been proposed, the most promising being the optical vortex phase mask coronagraphs. 
In this paper, a new scheme of broadband optical scalar vortex coronagraph is proposed and characterized experimentally in the laboratory.
Our setup employs a pair of computer generated phase gratings (one of them containing a singularity) to control the chromatic dispersion of phase plates and 
achieves a constant peak-to-peak attenuation below $1\cdot10^{-3}$ over a bandwidth of 120 nm centered at 700 nm. 
An inner working angle of $\sim \lambda/D$ is demonstrated along with a raw contrast of 11.5 magnitudes at $2\lambda/D$.
\end{abstract}



\section{Introduction}
Direct imaging of exoplanets \citep{Marois2008,Serabyn2010,Ralf2012} is one of the major scientific drivers behind the recent developments in stellar coronagraphy \citep{Mawet2012}. The possibility to distinguish the faint light of the planet amid the glare of the host star would eventually enable spectrometric characterization of their atmospheres \citep{Janson2010}, which is currently mostly possible for large size transiting exoplanets \citep{Charbonneau2002}. In this regard, the thriving challenge is the detection of biomarkers in the exoplanet atmospheres \citep{Lovelock1965}, and thus to ascertain the frequency of life in the universe. 

In traditional stellar coronagraphy \citep{Vilas1987}, the light of the star is blocked in the image plane by means of an opaque screen and then filtered in the pupil plane by a second opaque mask (the Lyot stop, \citet{Lyot1939}), before the image is formed on the detector. 
This configuration gives good suppression of the star light, at the expense of the inner working angle, \textit{i.e.} the 
minimal distance from the attenuated star at which a faint neighboring object can be observed.
Recently, attention was driven to the possibility to use phase rather than amplitude masks to suppress efficiently the star light while retaining an inner working angle close to the diffraction limit \citep{Roddier1997}.
Many different types of phase mask coronagraphs have been suggested and experimented in the last decade \citep{Mawet2012}, the common denominator being that the phase mask is used to scatter the light of the central star outside the pupil aperture and thus be easily removed by a circular aperture used as Lyot stop. Particularly interesting in this regard are the so called vortex phase mask coronagraphs \citep{Foo2005}, which transform the starlight beam into an optical vortex, a ring shaped beam with a null in the center and a spiral wavefront \citep{Allen1992}.  

Most of the proposed methods however suffer from strong chromaticism, due to the fact that a specific phase shift cannot be generated by a single mask for all wavelengths simultaneously \citep{Swartzlander2005}. As a result, phase mask coronagraphs can attain high rejection of the central star only if operated with small bandwidths, certainly a limitation for the goal of spectroscopic investigation of exoplanets.  

A solutions to the problem of chromaticism has been proposed by \citet{Mawet2005} and consists in generating a polarization optical vortex by exploiting the form birefringence of sub-wavelength gratings (so called vectorial vortex coronagraphs). While this method is possible for mid-infrared wavelengths \citep{Delacroix2013,Mawet2013}, for shorter wavelengths the nanotechnology for the development of the phase masks is not yet at the level required by the scaling. 
Vectorial vortex coronagraphs at shorter wavelengths were fabricated by means of liquid crystal polymers technology \citep{Mawet2009} and gave excellent broadband contrasts in the laboratory \citep{Mawet2011} and on sky \citep{Serabyn2010}.

In this paper, we propose a solution for the chromaticity of scalar optical vortex coronagraphy in the visible band by use of a method previously applied to femtosecond laser optics to generate broadband optical vortices. 
The peak-to-peak attenuation of our coronagraphic setup is constant below the $1\cdot10^{-3}$ level over a bandwidth of at least 120 nm centered at the wavelength of 700 nm. The throughput of the vortex generating set-up achieves a peak value of 75\%, and is approximately constant for wavelengths between 640 nm and 800 nm.

\section{Generation of broadband optical vortices} 

\subsection{Optical vortices and coronagraphy}
An optical vortex is an optical field containing a point where the phase is undefined, usually referred to as the \textit{phase singularity}. In this point the optical field has exactly 0 amplitude. The circulation of the phase transverse gradient around a path $\gamma$ around a phase singularity is an integer multiple $m$ of 2$\pi$, and is addressed as the \textit{topological charge} of the optical vortex:
\begin{equation}
\oint_\gamma \vec\nabla \phi(x,y)\cdot d\vec l = m 2\pi
\end{equation}
Here $\phi(x,y)$ is the phase in the transverse plane perpendicular to the propagation direction of the field. For a field $U_V$ with cylindrical symmetry we may write the optical vortex in the following form:
\begin{eqnarray}   
U_V(r,\theta,z)=&A( r ) \exp(ik_0z)\exp[-i\phi(\theta)]\nonumber\\
=&A( r ) \exp(ik_0z)\exp(-i m\theta)
\end{eqnarray}
Where $A( r )$ is a radial amplitude function with a null in the origin $r=0$. The wavefront of the vortex field (locus of points with the same phase) has the shape of a screw of step $2\pi/k_0$.
Phase singularities can be written on a generic field $A( r )$ by means of a phase mask, i.e. a plate where the local optical path is given by $m\theta/k_0$.
Particularly interesting for astronomical applications is the case in which $m=2$ and $A( r )=J_1( r )/r$, such as in the case 
of an aberration-free telescope with circular pupil focusing starlight on the center of the phase plate.
It can be demonstrated \citep{Mawet2005,Swartzlander2009} that in the pupil plane the field $U_V$ is transformed in the following form:
\begin{eqnarray}
U_P(r,\theta)={FT}\{U_V\}=\exp(-i2\theta)
\begin{cases}
0&  r<R_P\\
(R_P/r)^2&  r\geq R_P.
\end{cases}
 \end{eqnarray}
The equation shows that the light is scattered outside the radius $R_P$ of the pupil (the so called \textit{ring of fire}). 
This property, which is shared by vortices with even topological charge \citep{Mawet2005}, can be exploited to dim the light of the star efficiently. Indeed by placing a circular aperture with radius $R<R_P$ in the pupil plane we can in principle reject the whole light of a central star, while retaining most of the light from a weak companion which, focused outside the vortex region of the phase plate, will be distributed over the area of the pupil. 
A final pupil transformation (Fourier transform) will allow to image just the companion. In practice, the ring of fire is smoothed by aberrations and some light of the central star is scattered within the pupil. 

Another practical limitation of this scheme is that phase plates work only for a fixed wavelength \citep{Swartzlander2005} and operation in polychromatic regime degrades considerably the contrast of a phase plate coronagraph \citep{BESSEL}.
As we shall see in the next paragraph, a pair of holographic phase plates can resolve the problem.

\subsection{Off-axis vortex holographic phase plates}
The effect of the chromaticity of on-axis vortex phase plates gives rise to the superposition of vortices of various topological charges in the field transmitted by the phase plate \citep{Swartzlander2005}. 
This problem is solved by using an off-axis phase plate hologram.

A hologram is a recording of an optical field \citep{Goodman:1968}. By illuminating the hologram with an appropriate reference wave, the recorded field can be reconstructed and used for various applications. 
The hologram can be encoded in amplitude or phase.
In many applications, the hologram is generated numerically and transferred with photolithographic techniques to an optical substrate. We thus speak of computer-generated holograms (CGH). 
In phase holograms, local variations of the thickness or refractive index of a transparent substrate are used to encode the phase of an optical field.  

An off-axis vortex phase CGH can be generated by extracting the phase of the product between a phase singularity ($U(\theta)=\exp(-i m\theta)$) and a reference plane wave $U_R=\exp(-i \alpha k_0 x)$ propagating at an angle $\alpha$ from the optical axis.
The spatially resolved phase delay of the plate for $m=2$ is illustrated in Figure~1.a.
We see that the hologram is akin to a diffraction grating with a period $g=2\pi/(\alpha k_0)$ and a  phase singularity in the center (three lines converge into one).
By illuminating the hologram with an optical field, we will in general scatter the light over different diffraction orders of the grating. The field scattered at diffraction order $n$ is a vortex with a topological charge $n\cdot m$. At the design wavelength, the phase hologram will be scattering the light in the first order of diffraction only. When the phase plate is illuminated by light at a different wavelength, light will be scattered also to other diffraction orders, but preferentially to the first order. Therefore, the superimposed grating will sort spatially the contribution of the parasitic topological charges which represent the problem of the on-axis phase plate \citep{Swartzlander2005}.  

\begin{figure}
\includegraphics[width=9cm]{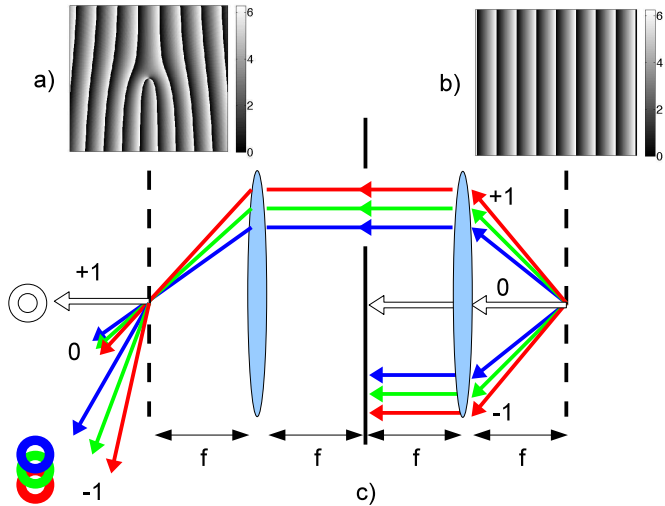}
\caption{\label{fig:vortex-scheme} Cartoon illustrating the operation of the broadband scalar vortex generator. a) Gray scaled image of the computer generated phase hologram for the generation of optical vortices with topological charge $m=2$. The gray scale represents the phase in radians. b) Gray scaled image of the phase grating used to disperse angularly the incoming light from the telescope. c) The scheme of the broadband scalar vortex generator. Light propagates from right to left. After being diffracted by the phase grating (see insert b), the first diffraction order is selected in the back focal plane of a lens (focal length f) by means of an aperture. The dispersed light is then refocused by a second lens onto the phase singularity of the computer generated hologram (see insert a). The beam scattered at the first order of diffraction is not angularly dispersed and contains a vortex of charge $m=2$ in its center.}
\end{figure}

However, the diffraction angle at the first order will be proportional to the wavelength. Therefore, by illuminating the hologram with white light we will obtain a superposition of monochromatic copies of the optical vortex, each propagating at a slightly different direction according to its wavelength. This is a manifestation of the angular dispersion introduced by the grating. 
A way to overcome this problem is to illuminate the hologram with angularly dispersed light propagating along the direction of its first diffraction order. Light will be mainly scattered at order +1 of the vortex hologram and the angular dispersion will be cancelled leaving a white light optical vortex beam.
A practical implementation of the scheme is illustrated in Figure~\ref{fig:vortex-scheme}.c (see also \citet{PaulusOL}). A first grating (Fig.~1.b) introduces angular dispersion in the white light beam. The first diffraction order is selected by a lens and a slit aperture and re-imaged by a second lens onto the vortex hologram, which exhibits a grating with the same periodicity as the first grating.  
The vortex hologram will add the topological charge to the beam and for the diffraction order +1 will also recompose the degeneracy of the propagation directions of the various colors, so that a white light vortex is formed.
Notice that all defects of the vortex induced by the chromaticity of the phase plate will propagate at different diffraction orders which will be rejected. Thus the inability of the vortex phase plate in generating pure charge vortices for broad bandwidth will be transferred to a throughput inefficiency of the setup. 

\section{Experimental setup}

\begin{figure}
 \includegraphics[width=9cm]{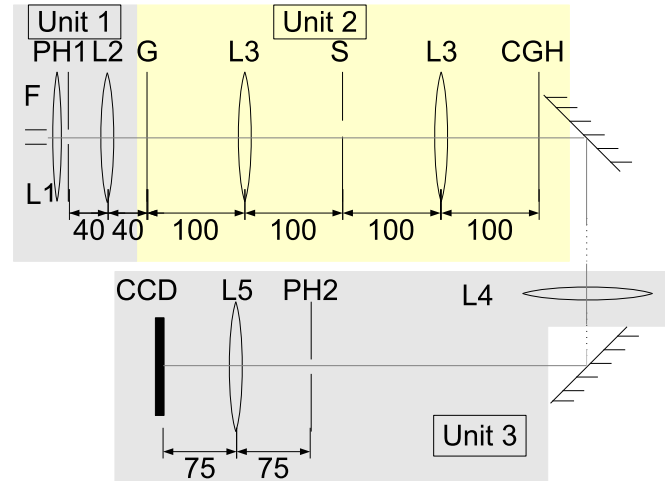}
 \caption{Our arranged setup with two mirrors to fold the path of light. The lenses L1 ($f=11$\,mm) and L2 ($f=40$\,mm) and the pinhole PH1 ($D=75\,\mu$m) create the Airy-Disk from the Fiber F. All but the first diffraction order from grating G are shielded (S). This order is imaged with L3($f=100$\,mm) to the Hologram (CGH) and the vortex beam is re-imaged to a Lyot stop PH2 ($d=700\,\mu$m), where the ring of fire is blocked with a pinhole. The image of PH2 is formed with L5 ($f=75$\,mm) to the CCD.}
 \label{fig:setup}
\end{figure}

The setup for our laboratory test of the broadband optical vortex coronagraph is composed by three main functional units and is depicted  in Figure~\ref{fig:setup}. The first functional unit is used to simulate the diffraction pattern of an obstruction-free telescope. 
We used light either from a HeNe laser ($\lambda_0=633$\,nm) or from a broadband white light source ($\lambda_0\ge640$\,nm) coupled into a single mode fiber to simulate a point-like source. The light from the optical fiber was collimated by a $f=11$\,mm aspheric lens (L1) and used to illuminate a pinhole (PH1) of diameter $D=75\,\mu$m which simulates the pupil of a telescope. A $f=40$\,mm lens (L2) was used to generate an Airy disk with radius $r=420\,\mu$m (first minimum radius, $\lambda=640$\,nm) at the input of the second functional unit (broadband optical vortex generator).

The scheme of the broadband optical vortex generator is the replica of that illustrated in Figure~\ref{fig:vortex-scheme} implemented with $f=100$\,mm lenses and blazed phase diffraction gratings with period $g=50\,\mu$m. The gratings (Fraunhofer Institute, Jena) were prepared by replicating on a thin polymer layer (10 $\mu$m) a master hologram obtained by gray tone laser lithography. They are optimized for the wavelength of $\lambda_0=650$\,nm and have a maximum diffraction efficiency in the first order exceeding 90\%. 
The vortex hologram (CGH) was placed in the focus of the L3 on a x-y micrometric translation stage which allowed us to align precisely the phase singularity of the grating with the optical axis of the setup.
An iris blind was used to select just the first diffraction order of the first grating. 

In the third functional unit (Lyot stop), the image plane is transformed back to the pupil plane by a $f=400$\,mm achromatic lens (L4). A ring of fire with inner radius of $350\,\mu$m and maximum intensity at $410\,\mu$m was formed in the back focal plane of the lens. In this plane we placed a $d=700\,\mu$m pinhole which corresponds to 93\% of the magnified diameter of the entrance pupil ($750\,\mu$m).
Finally, lens L5 transformed back the optical field into the image plane, which was recorded with an 8-bit CCD camera. 

To achieve high dynamic range with the CCD, we exploited the 12-bit dynamics of the shutter speed. 
High dynamic range images are obtained by 1) taking several shots with different shutter speed, 2) removing the saturated pixels, 3) applying the dark correction, 4) scaling the pixel values to the shutter speed, and 5) combining the resulting partial images by the weighted average. Thereby the combination is done pixel by pixel and the flux after step 3) is used for the weighting, hence we avoid the upscaling of dark current noise because of the low weight of faint pixels. 
The linearity of both the pixel and shutter speed dynamics were tested. The estimated overall dynamics of our images is the product of the dynamic of the camera with the dynamic of the shutter, thereby $2^{20}\approx10^{6}$.

\section{Results}
To gauge the performance of the coronagraph, we measured first the peak-to-peak attenuation, defined as the ratio between the maximum flux in the on-axis radial profile and the maximum flux in the off-axis radial profile. To this end, we built two high dynamic images of the final focal plane with 1) the phase singularity of the CGH centered on the optical axis of the instrument and 2) with the phase singularity shifted off-axis by several $\lambda/D$. In the first case, the ring-of-fire of the artificial star will be blocked by PH2, while in the second case, the light of the artificial star will be transmitted through PH2. 

\begin{figure}
 \includegraphics[width=9cm]{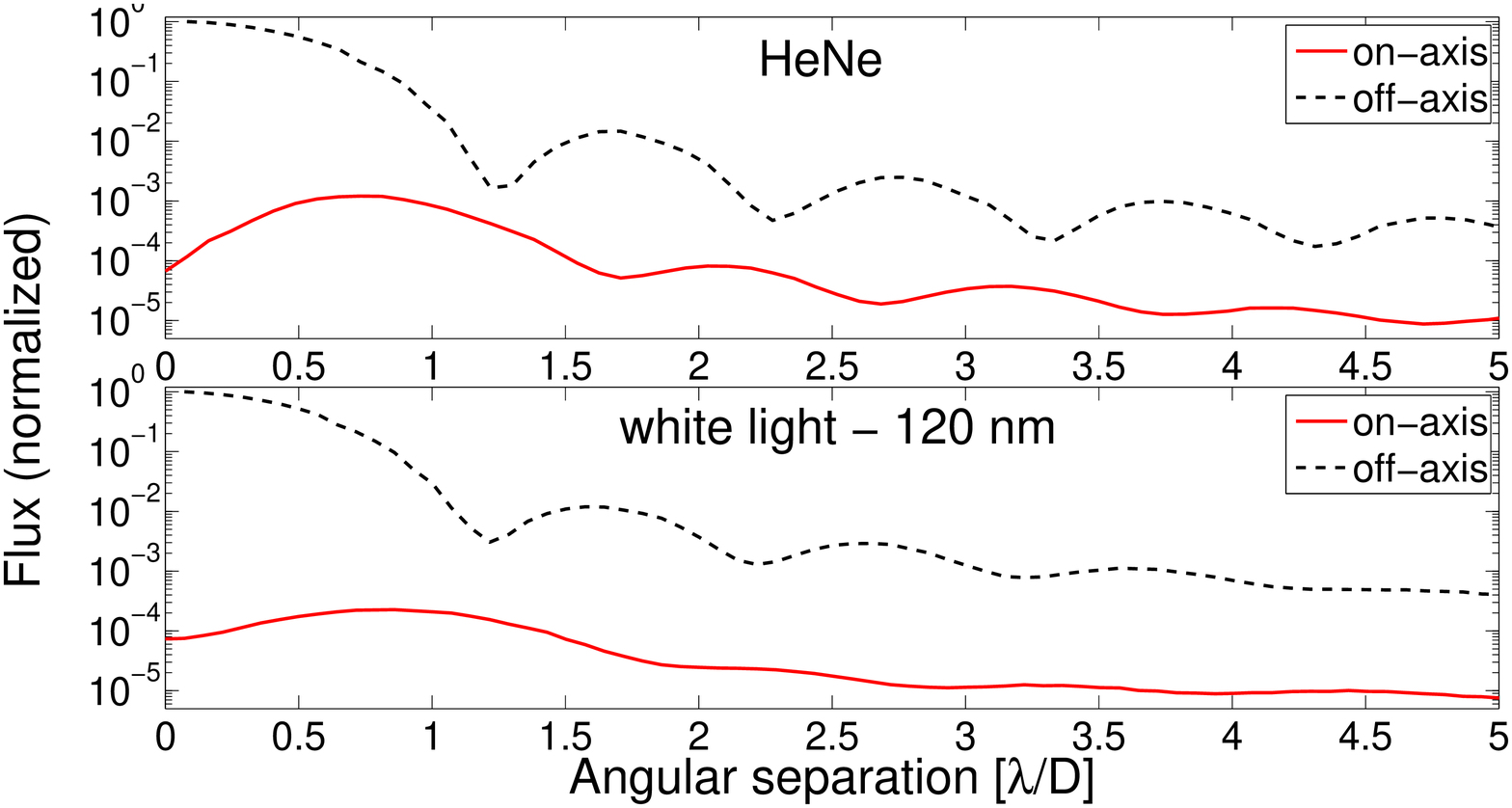}	
 \caption{Radial profiles for HeNe laser and broad band light ($\Delta\lambda=$120\,nm, centered at $\lambda=$700\,nm). The brightness progress is shown for the CGH on-axis (red continuous line) and for the CGH off-axis by $>3\lambda$/D(black dashed line). The difference between the maximum fluxes is about 3 orders of magnitude.}
 \label{fig:mitohnevortex}
\end{figure}

Figure~\ref{fig:mitohnevortex} shows the azimuthally integrated profiles of these two images for an artificial star created with a HeNe laser and a $\Delta\lambda=$120\,nm broadband light source. The broadband light source was achieved by setting 7 equally spaced channels covering the wavelength range from 640 to 760\,nm in our programmable white light source. 
For both light sources a peak-to-peak attenuation to $\sim 5\cdot10^{-4}$ was observed, showing the intrinsic achromatic behavior of our coronagraph. In the broadband case the attenuation at $\lambda/D$ converts to 9.2\,mag, while the attenuation at 2$\lambda/D$  reaches a level of $2.4\cdot10^{-5}$ (11.5\,mag), a performance comparable to the laboratory tests of the vectorial vortex coronagraph \citep{Delacroix2013}. 

\begin{figure}
 \includegraphics[width=9cm]{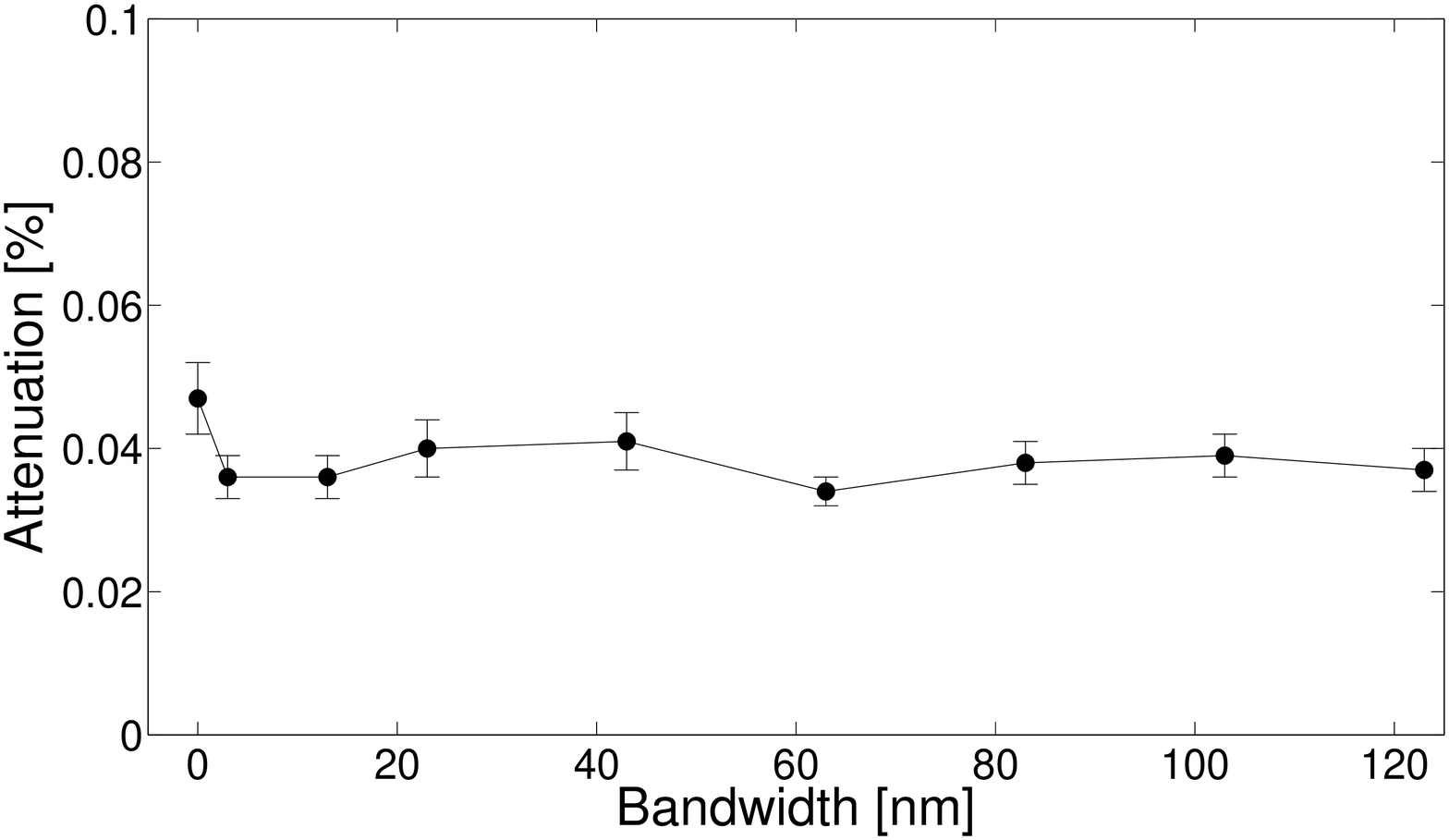} 
 \caption{The peak-to-peak attenuation for different bandwidths of the light. The attenuation is constant over the covered bandwidth of 120\,nm, to which we are limited by our instruments.}
 \label{fig:contrast}
\end{figure}

The dependence of the peak-to-peak attenuation on the bandwidth of light centered at 700\,nm is shown in Figure~\ref{fig:contrast}. As before, the bandwidth was achieved by selecting appropriately the 8 channels available at our white light source in steps of 10 or 20\,nm. 
In the covered range of bandwidths, the peak-to-peak attenuation is constant within the measurement errors to a value of ($0.037\pm0.004$)\%. At the maximum bandwidth (120 nm), the relative bandwidth $\Delta\lambda/\lambda$ is 17\%, enough to cover the band of an astronomical filter.
We notice however that our experiments were not able to investigate the maximal bandwidth of our coronagraphic setup since we were limited by the wavelength range of the white light source (emitting for wavelengths $\lambda\ge640$\,nm, but efficiency drops for $\lambda>780$\,nm).
There is no reason that the peak-to-peak attenuation drops for even bigger bandwidths. The only limit is given when the different orders of the first grating begin to overlap. For the wavelength to overlap between the first and second order of diffraction we need to have wavelengths  $\lambda<434$\,nm and $\lambda>870$\,nm, thus allowing a  spectral range of $\sim430$ nm centered at 650 nm. This remarkable bandwidth range will include the V and R bands while covering half of the B band and most of the I band.

\begin{figure}
 \includegraphics[width=9cm]{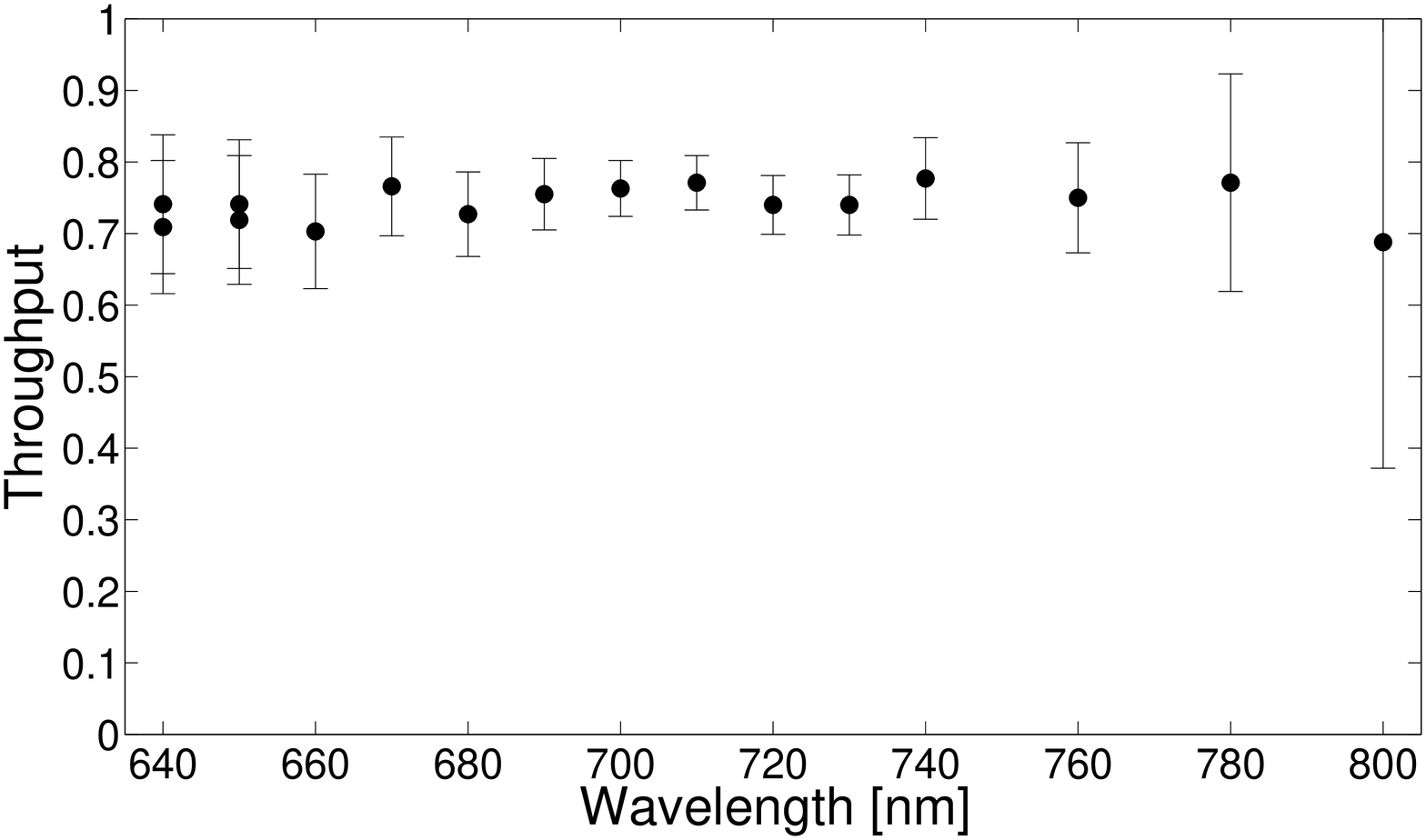}	
 \caption{Transmission of the vortex generation unit (Unit~2 in Figure~\ref{fig:setup}). The values for the wavelength of 640\,nm and 650\,nm were measured twice to check stability of the measurements. The efficiency is constant over the wavelength range 640-800\,nm. The measurements were limited by the used instruments, in principle, operation at wavelength outside the covered interval would be possible.}
 \label{fig:transmission}
\end{figure}

A potential limitation of our setup is however that the blazing of the grating and CGH may loose diffraction efficiency for wavelengths strongly detuned from the design wavelength. We show however that this is not the case for the investigated wavelength range. 
Figure~\ref{fig:transmission} shows the throughput of the vortex generator functional unit, \textit{i.e.} comprising the optical components  from the grating to the hologram (see Figure~\ref{fig:vortex-scheme}). The weighted average transmission is 75\%, and it is constant over the wavelength range of 160\,nm. Longer wavelengths could not be explored due to low efficiency of the white light source and detector.
Notice that the grating and hologram were not coated with anti-reflection layer, hence reflexion losses are expected to occur at these surfaces. Assuming a conservative reflectivity of 4\% at each surface, losses of $\sim$16\% occur, meaning we could push the throughput of the whole system to 90\% by using anti-reflection coating.

\begin{figure}
 \includegraphics[width=9cm]{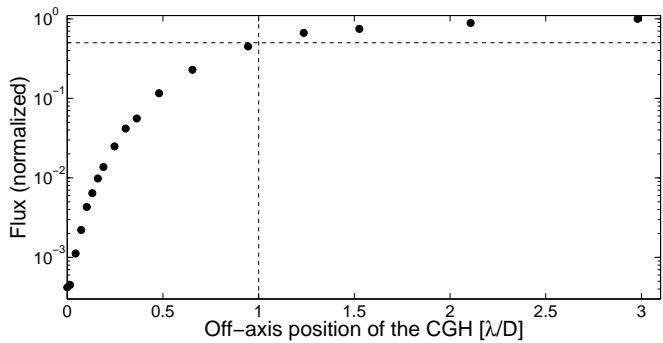}	
 \caption{The flux of the brightest pixel as function of the angular separation between beam and phase singularity in the CGH, normalized to the flux at separation $3 \lambda/D$. The dashed lines indicate attenuation of 50\% and separation of 1$\lambda/D$.}
 \label{fig:offaxisflux}
\end{figure}

Finally, we simulate the effect of the CGH on a close companion by measuring the transmission of our setup while moving the CGH off axis. Figure~\ref{fig:offaxisflux} shows the normalized transmission of the brightest spot plotted against the off-axis position of the CGH. At an angular separation $\lambda/D$ about 50\% of the light is suppressed while a companion at separation $\ge2 \lambda/D$ is nearly unaffected by the phase singularity in the CGH.
This result is in line with the expectations of the narrow inner working angle of the vortex coronagraph. The data also show that the inherent stability of the angular separation between the beam and the CGH singularity should be maintained below 0.044 $\lambda/D$ in order to ensure a contrast better than $10^{-3}$.

\section{Conclusions and perspectives}

In conclusion, we have performed the laboratory characterization of a broadband optical vortex coronagraph. 
We showed, that a high contrast over large bandwidth is reachable, which is a key point to do spectroscopy on faint companions.

Compared to other schemes of scalar vortex coronagraphy \citep{BESSEL}, the proposed setup has the advantage of 
providing very high attenuations of the central star over much larger bandwidths. We experimentally verified that the contrast can be kept constant over bandwidths of 120 nm centered at the wavelength of 700 nm, enough to cover a transmission bandwidth of the atmosphere. As mentioned, however, the setup can in principle provide high peak-to-peak attenuations of the central star also for larger bandwidths albeit with a lower throughput.
We note that the laboratory performance of our set-up is comparable to that of the vectorial optical vortex coronagraph \citep{Delacroix2013} in terms of relative bandwidth and peak-to peak attenuation. Indeed our setup exhibits a contrast nearly 10 times better than the vectorial vortex coronagraph. 
In principle, our setup can be extended to the near- or mid-infrared, where for both young and old exoplanets the brightness difference to their host stars is much smaller than in the visible. Assuming that we can achieve the same efficiency of the broadband scalar vortex coronagraph in the mid-infrared, we notice that a peak-to-peak attenuation below $\approx 0.1$\% in combination with a camera dynamics of 10$^4$ and image processing techniques \citep{LOCI} would allow to detect objects as faint as 10$^{-6}-10^{-7}$ ($\Delta m=15-17$ magnitudes) compared to the central star, thus permitting to image directly old exoplanets. 

As for the other vortex coronagraphs based on the generation of charge 2 vortices, the inner working angle has been shown to be $\lambda/D$, while the field of view will be limited only by the field of view of the initial image plane.  
Notice that our scheme is suitable also for the generation of vortices with topological charge greater than 2. As known \citep{Guyon2007,Jenkins2008}, higher order vortex coronagraphs are less sensitive to low order aberrations (\textit{e.g.} tip-tilt) and intrinsic stellar disk size, the trade-off being represented by larger inner working angles. 
In case of CGH, the resolution of the fabrication method (in our case 1 $\mu$m) will be the limiting factor for the realization of holograms with topological charge larger than 2, since the definition required to write a singular point increases with the topological charge of the vortex. Work is in progress to assess the impact of the resolution of the phase mask on the contrast.    

The price to pay for a large operation bandwidth of the proposed coronagraph is the fact that the imprint of the vortex on the optical wavefront requires a relatively complex setup as compared to the single mask approach. The presence of two lenses and alignment degrees of freedom may introduce additional scattering sources and aberrations that may limit the achievement of even higher attenuations than already demonstrated. 
It may be worth noting that a simplified setup for a similar broadband vortex generation was proposed and tested by \citet{Mariyenko2005}. 

We believe our result represent a significant step towards the realization of high performance, broadband coronagraphs for the next generation telescopes equipped with extreme adaptive optic systems which will enable routine imaging and spectroscopy of exoplanets and other faint features around stars.

\section*{Acknowledgments}
We thank the referee, Dimitri Mawet, for careful and helpful reading of this paper.


\begin{thebibliography}{99}
\bibitem[\protect\citeauthoryear{Allen et al.}{2004}]{Allen1992} Allen L., Beijersbergen M. W., Spreeuw R. J. C., Woerdman J. P., 1992, Phys. Rev. A, 45, 8185
\bibitem[\protect\citeauthoryear{Bezuhanov et al.}{2004}]{PaulusOL} Bezuhanov K., Dreischuh A., Paulus G.G., Sch\"atzel M. G.,  Walther H., 2004, Opt. Lett., 29, 1942
\bibitem[\protect\citeauthoryear{Charbonneau et al.}{2002}]{Charbonneau2002} Charbonneau D., Brown T. M., Noyes R. W., Gilliland R. L., 2002, ApJ, 568, 377
\bibitem[\protect\citeauthoryear{Delacroix et al.}{2013}]{Delacroix2013} Delacoix C. et al., 2013, A\&A, 553, A98
\bibitem[\protect\citeauthoryear{Foo, Palacios \& Swartzlander}{2005}]{Foo2005} Foo G., Palacios D.M., Swartzlander G.A., 2005, Opt.Lett., 30, 3308
\bibitem[\protect\citeauthoryear{Goodman}{2005}]{Goodman:1968} Goodman J. W., 2005, "Introduction to Fourier optics", Roberts \&Co. Publishers, Greenwood Village.
\bibitem[\protect\citeauthoryear{Guyon}{2007}]{Guyon2007} Guyon O., 2007, C. R. Phys., 8, 323 
\bibitem[\protect\citeauthoryear{Janson et al.}{2010}]{Janson2010} Janson M., Bergfors C., Goto M., Brandner W., Lafreni\`ere D., 2010, ApJ, 710, L35
\bibitem[\protect\citeauthoryear{Jenkins}{2008}]{Jenkins2008} Jenkins C., 2008, MNRAS, 384, 515
\bibitem[\protect\citeauthoryear{Lafreni\`ere et al.}{2006}]{LOCI} Lafreni\`ere D., Marois C., Doyon R., Nadeau D., Artigau E. 2006, ApJ, 660, 770
\bibitem[\protect\citeauthoryear{Lovelock}{1965}]{Lovelock1965} Lovelock J.~E., 1965, Nature, 207, 568
\bibitem[\protect\citeauthoryear{Lyot}{1939}]{Lyot1939} Lyot B., 1939, MNRAS, 99, 580
\bibitem[\protect\citeauthoryear{Mariyenko, Strohaber \& Uiterwaal}{2005}]{Mariyenko2005} Mariyenko I. G., Strohaber J., Uiterwaal C.J.G.J., 2005, Opt. Exp., 13, 7599 
\bibitem[\protect\citeauthoryear{Marois et al.}{2008}]{Marois2008} Marois C., Macintosh B., Barman T.,  Zuckerman B., Song I., Patience J., Lafreni\`ere D., Doyon R., 2008, Science, 322, 1348
\bibitem[\protect\citeauthoryear{Mawet et al.}{2005}]{Mawet2005} Mawet D., Riaud P., Absil O., Surdej J., 2005,
Appl. Opt., 44, 7313
\bibitem[\protect\citeauthoryear{Mawet et al.}{2009}]{Mawet2009} Mawet D., Serabyn E., Liewer K., Hanot Ch.,  McEldowney S.,
 Shemo D., OÕBrien N., 2009, Opt. Exp., 17, 1902
\bibitem[\protect\citeauthoryear{Mawet et al.}{2011}]{Mawet2011} Mawet D., et al., 2011,
Proc. of SPIE, 8151, 8151D1
\bibitem[\protect\citeauthoryear{Mawet et al.}{2012}]{Mawet2012} Mawet D., et al., 2012,
Proc. of SPIE, 8442, 844204
\bibitem[\protect\citeauthoryear{Mawet et al.}{2013}]{Mawet2013} Mawet D., et al., 2013, A\&A, 512, L13
\bibitem[\protect\citeauthoryear{Neuh\"auser \& Schmidt}{2012}]{Ralf2012} Neuh\"auser R., Schmidt T.O.B., 2012,
`Direct Imaging of extra-solar planets - homogeneous comparison of detected planets and candidates' in Tyson R.K. (Ed.) {\it Topics in Adaptive Optics},
astro-ph:1201.3537
\bibitem[\protect\citeauthoryear{Roddier \& Roddier}{1997}]{Roddier1997} Roddier F., Roddier C.,1997,
PASP, 109, 815
\bibitem[\protect\citeauthoryear{Serabyn, Mawet \& Burruss}{2010}]{Serabyn2010} Serabyn E., Mawet, D. , Burruss R., 2010, Nature, 464, 1018
\bibitem[\protect\citeauthoryear{Swartzlander}{2005}]{Swartzlander2005} Swartzlander G. A. Jr., 2005, Opt. Lett., 30, 2876
\bibitem[\protect\citeauthoryear{Swartzlander et al.}{2008}]{BESSEL} Swartzlander G. A. Jr., Ford E. L., Abdul-Mailk R. S.,
Close L. M., Peters M. A., Palacios D. M., Wilson D. W., 2008, Opt. Exp., 16, 10200
\bibitem[\protect\citeauthoryear{Swartzlander}{2009}]{Swartzlander2009} Swartzlander G. A. Jr.,  2009, J. Opt. A, 11, 094022
\bibitem[\protect\citeauthoryear{Vilas \& Smith}{1987}]{Vilas1987} Vilas F., Smith B. A., 1987, Appl. Opt., 26, 664
\end{thebibliography}
\end{document}